\documentstyle[aps,preprint]{revtex}
\begin{document}
\title{Enhanced winnings in a  mixed-ability population \\
playing a minority game}
\author{N.F. Johnson$^{a}$, P.M. Hui$^{b}$, Dafang
Zheng$^{c}$, and M. Hart$^{a}$}
\address{$^{a}$ Physics Department, Clarendon Laboratory,
Oxford University, Oxford OX1 3PU, U.K.}
\address{$^{b}$ Department of Physics, The Chinese University
of Hong Kong, Shatin, \\ New Territories, Hong Kong}
\address{$^{c}$ Department of Applied Physics,  South China
University of Technology, 
\\ Guangzhou 510641, P.R. China}
\maketitle
\begin{abstract} 
We study a mixed population of adaptive
agents with small and large  memories, competing  in a
minority game. If the agents are sufficiently adaptive, we
find that the average winnings per agent can exceed that
obtainable in the corresponding pure populations. In contrast
to the pure population, the average success rate of the
large-memory agents can be greater than 50 percent. The
present results are not reproduced if the agents are fed a
random history, thereby demonstrating the importance of memory
in this system.  

\vspace*{0,5 true in} 

\vspace*{0.2 true in}
\end{abstract}

\newpage

The subject of emergent phenomena in complex adaptive systems
is attracting much interest among workers in  physics,
economics, biological and social sciences
\cite{holland}. The minority game, introduced by Challet and
Zhang
\cite{challet} and studied subsequently by various
workers\cite{savit,johnson2,decara,cavagna,rodgers},
represents a fascinating toy-model of a dynamical, complex
adaptive system where members of the population (agents)
repeatedly compete to be in a minority. It offers, for
example, a simple paradigm for the decision dynamics
underlying financial markets: more buyers than sellers implies
higher prices, hence it can be better for a trader to be in
the minority group of sellers.  Subsequent to Challet and
Zhang's pioneering work \cite{challet}, studies of the
minority game have focused on a `pure' population where all
agents have the same memory, or brain-size,
$m$. A particularly interesting observation by Savit
\cite{savit} is that while correlations can arise in the
resulting (endogenous) history time-series for the pure
population, these correlations cannot be further exploited for
profit by the agents themselves. 

In this paper, we explore the performance of a mixed-ability
population where individual agents have either a small or
large  memory. If the agents are sufficiently adaptive, we
find that the average winnings per agent can exceed that
obtainable in the corresponding pure populations, achieving a
maximum at a non-zero mixing ratio. In contrast to the pure
population, the average success rate of the large-memory
agents can exceed
$50\%$. These results, however, are {\em not} reproduced if
the agents are fed a random history, thereby demonstrating the
importance of the true history (and hence memory) in this
system.

The minority game
\cite{challet,savit,johnson2,decara,cavagna,rodgers}  takes
the form of a repeated game with an odd number of agents $N$
who must choose independently whether to be in room `0' or
room `1'.   The winners are those in the room with fewer
agents. The `output' is a single binary digit, 0 or 1,
representing the winning room for each time step.  This output
is made available to all agents, and is the {\em only}
information they can use to make decisions in subsequent
turns.   The memory $m$ is the length of the recent history
bit-string that an agent can use when making its next decision
\cite{challet,savit,johnson2,decara,rodgers}. We consider a 
population containing
$n_m$ agents with memory $m$, and $n_{m'}=N-n_m$ agents with
memory $m'$.  Consider
$m=3$;  there are
$2^{2^m}=256$ possible strategies, each of which can be
represented by a string of 8 bits ($0$ or $1$) corresponding
to the decisions based on the $2^3 = 8$ possible histories
000, 001 etc. . The agents randomly pick
$s$ strategies at the beginning of the game, with repetitions
allowed. After each turn, the agent assigns one (virtual)
point to each of his strategies which would have predicted the
correct outcome. In addition the agent gets awarded one (real)
point if he is successful. At each turn of the game, the agent
uses the most successful strategy (i.e. most virtual points) 
from his bag of
$s$ strategies.   The success of any particular strategy is
generally short-lived. If all the agents begin to use similar
strategies,  such a strategy ceases to be profitable and is
dropped. Hence there is no best strategy for all times. 

There are two `spaces' of interest in the minority game, both
of which depend on $m$:

\noindent (a) {\em Strategy space}. This forms a
$2^m$-dimensional hypercube for memory $m$ with strategies at
the $2^{2^m}$ vertices \cite{challet}. If the size of the
strategy space is small  compared to the total number of
agents $N$ (i.e. $2.2^m<<Ns$
\cite{challet,savit,johnson2}) many agents may hold the
highest-scoring strategy at any given turn and hence make the
same decision. This leads to a large standard deviation in the
winning room and hence a relatively low number of total points
awarded
\cite{challet,savit,johnson2,decara,rodgers}.  Such
crowd-effects are a strategy-space phenomenon and have been
shown to quantitatively explain the fluctuations for the pure
population as a function of $m$ and
$s$ \cite{johnson2}.  Consider the average points per agent
per turn $W$, defined as the total number of points awarded in
that turn divided by the total number of agents.   For small
$m$, $W$ is substantially less than 0.5 due to the
crowd-effects mentioned above. The maximum possible 
$W$ would correspond to the winning-room attendance
permanently staying at
$(N-1)/2$. Therefore 
$W$ is always less than or equal to $(N-1)/2N$, hence
$W<0.5$.  Note that an external (i.e. non-participating)
gambler using a coin-toss to predict the winning room, would
have a
$50\%$ success rate since he would not suffer from this
intrinsic crowding in strategy-space.

\noindent (b) {\em History space}. This forms an 
$m$-dimensional hypercube whose $2^m$ vertices correspond to
all  possible recent history bit-strings of length $m$ (000,
001 etc. for $m=3$). For a pure population of agents with
memory $m$, where $2.2^m<<Ns$, there is information left in
the history time-series
\cite{savit}: however this information is hidden in
bit-strings of length greater than
$m$ and hence is not accessible to the agents \cite{savit}.
For large
$m$, there is information left in bit-strings of any length:
however the agents have insufficient strategies to further
exploit this information \cite{savit}.  Cavagna has shown
\cite{cavagna} that the standard deviation of the attendance
of a given room (e.g. room 0) is changed little if the real
history is replaced by a random one at each turn of the game:
this follows from the fact that the agents cannot exploit any
remaining  correlations in the real history time-series. 

Figure 1 shows the average winnings per agent per turn $W$ for
a mixed population of 
$N=101$ players, with memory $m=3$ or $m'=6$, obtained using
numerical simulations. The number of
$m=3$ agents
$n_3$ is shown on the x-axis, hence the number of $m'=6$
agents is given by $n_6=101-n_3$. Each agent has
$s=7$ strategies.  The parameters are chosen so that the pure
populations lie in the crowded, or frustrated, regime (i.e.
$2.2^m<<Ns$) since we are interested in investigating whether
our mixed population can adapt to relieve this frustration.
Each data-point corresponds to an average over 32 runs, with
data being collected in the limit of long times. For clarity
the spread in values at each data-point is not shown; these
spreads are sufficiently small for us to be certain that the
features we discuss here are not numerical artifacts. For the
`real history' results, the
$m=3$ ($m'=6$) agents are supplied with the last 3 (6) bits of
the winning-room time-series. For the `random history'
results, a random
$6$-bit string is generated at every turn of the game and
given to the $m'=6$ agents; the
$m=3$ agents are only given the last 3 digits of this string. 
A striking result of Fig. 1 is that the average number of
points per agent $W$ (solid circles) shows a {\em maximum} at
finite mixing, i.e. at $n_3 \sim 35$, $n_6\sim 65$. The total
number of points awarded per turn can therefore exceed that
corresponding to either pure $m=3$ or pure
$m'=6$ populations. This result is {\em not} reproduced if the
history supplied to the agents is random (open circles).

Figure 2 shows the average winnings per turn for the $m=3$ and
$m'=6$ agents separately. The straight line at 0.5 
corresponds to the average success-rate for a
non-participating agent predicting winning rooms using a
coin-toss; as discussed above, it is also the upper limit for
the average winnings per agent in a pure population. For
$n_3>45$, the average
$m'=6$ agent (solid triangles) manages to beat the coin-toss
agent and systematically profit from the game (see also Ref.
\cite{challet}). This cannot occur in the minority game with
pure populations, regardless of the memory
$m$. Although certain agents in the pure population may manage
to beat the coin-toss if the strategy space is sufficiently
large, the average winnings per agent is still less than 0.5.
It is surprising that this systematic profiteering by
$m'=6$ agents in the mixed population  can occur up to
relatively large values of $n_6$, i.e. in populations of up to
$\sim 55\%$
$m'=6$ players. Although the random-history result reproduces
the $m=3$ agents' winnings fairly well (open squares), it is
quite inaccurate for the $m'=6$ agents: specifically, the
$m'=6$ players in the random-history game (open triangles)
always do worse than the coin-toss. This is essentially
because the
$m'=6$ agents in the real-history game have the opportunity to
exploit correlations in the real-history time-series left by
the $m=3$ agents. The enhanced average winnings for
$m'=6$ agents in Fig. 2, together with the corresponding peak
in Fig. 1, result from the subtle interplay of history-space
and strategy-space effects and demonstrate the importance of
the memory in this system. 

Figure 2 shows that as $n_3 \rightarrow 1$, and hence
$n_6\rightarrow N-1$, the  remaining
$m=3$ agents manage to {\em increase} their average winnings.
In fact a single
$m=3$ agent in an $m'=6$ population wins the same amount as
the average $m'=6$ agent. This surprising result can be
understood by realizing that for  $n_6
\rightarrow N$, the
$m'=6$ population is in the `crowd' region
\cite{challet,savit,johnson2}, hence the
$m'=6$ agents win a relatively low number of points on
average. For
$n_3 >60$, the average winnings per $m=3$ agent saturates at
the value corresponding to a pure $m=3$ population, despite
the fact that the $m'=6$ agents are still systematically
profiting from the game. In this sense, our competitive agents
are also demonstrating some degree of (unintentional) mutual
cooperation.

As the number of strategies per agent $s$ decreases, for fixed
$m=3$ and $m'=6$, the peak in Fig. 1 becomes less pronounced.
For $s=2$, there is no discernible peak. This suggests that
the benefits of a mixed population, in terms of maximizing the
total number of points awarded, may require a certain level of
adaptivity. Even at $s=2$, however,  the average winnings of
the large-memory $m'=6$ agents is still greater than 0.5 in
the limit of  small
$n_6$.  Note that the main results presented here do not vary
significantly if $N$ is increased. Detailed results as a
function of $m$, $m'$, $s$ and
$N$ will be presented elsewhere. 

We now discuss the reasons why the real and random histories
might produce different results.  Consider the $2^{6}=64$
possible histories for
$m'=6$. Suppose the system has reached ...001010 at time $t$.
The subsequent evolution on the 6-dimensional history-space
hypercube allows a number of possible trajectories starting at
the vertex 001010. For example, at time
$t+1$ the system can be at 010100 or 010101; however it {\em
cannot} be at any other of the 64 possible vertices. We
therefore see that the $m'=6$ agents pass through the
history-space, and hence update their strategies, by moving on
a particular subset of all possible trajectories. In contrast,
the random-history model allows the system to jump to any
history-space vertex with equal probability at a given
time-step.   If the game only contains a pure population, the
agents cannot access any additional correlations in the
real-history time-series
\cite{savit}: it therefore matters little if the system visits
all history-space vertices in a deterministic or random way.
The random-history and real-history results are therefore
similar for pure populations (see Fig. 1 and Ref.
\cite{cavagna}). However when the two sets of agents coexist,
there are two separate time-scales for passing through the
respective history-spaces, i.e. 
$\tau_6>2^6=64$ for $m'=6$ agents and $\tau_3>2^3=8$ for $m=3$
agents.  The time-scale
$\tau_3$ for $m=3$ agents will be relatively short, hence
random-history and real-history models should provide similar
results for $m=3$ agents, as demonstrated in Fig. 2. On this
$\tau_3$ time-scale, however, the
$m'=6$ real-history agents are dynamically updating their
strategies in a  qualitatively different way from the
$m'=6$ random-history agents, because of this distinction
between real-history and random-history trajectories. The
$m'=6$ real-history agents  now have the opportunity to
exploit specific correlations in the real-history time-series
left by the $m=3$ agents.

Fluctuations in the winning room attendance away from
$(N-1)/2=50$ imply wasteage of total points \cite{challet}.
Hence $W \sim 0.5-\frac{\sigma}{N}$, where $\sigma$ is the
standard deviation of the attendance in one of the rooms, say
room 0. For a pure population in the crowded regime,
$\sigma
\propto N$: hence we write $\sigma \sim C N$ where $C$ is a
constant of proportionality (see Ref. \cite{johnson2} for a
more detailed analytic analysis). This expression includes the
crowd-effect in strategy-space involving agents with equal
memory \cite{johnson2}. For the mixed population, we assume
that the corresponding
$\sigma$ is obtained by adding separately the contributions to
the variance from the $m=3$ agents and the
$m'=6$ agents; defining the concentration of $m=3$ agents as
$x=\frac{n_3}{N}$, then
$\sigma^2 \sim
\sigma_3^2 + \sigma_6^2$ where $\sigma_3 \sim C_3 x N$ and
$\sigma_6 \sim C_6 (1-x) N$. This latter assumption of
additive variances assumes that the  system has managed to
remove any internal frustration between agents with different
memories, i.e. the two groups of agents behave independently.
This is not unreasonable given the 
interplay between history and strategy-spaces in the
real-history model. Hence in the mixed population $W\sim 0.5 -
[C_3^2 x^2 + C_6^2 (1-x)^2]^{1/2}$. For $x\rightarrow 0$ we
obtain a straight line $W\sim 0.5 - C_6 + C_6 x$ with
positive slope (consistent with real-history results in Fig.
1), while for
$x\rightarrow 1$ we obtain a straight line $W\sim 0.5 -
C_3 x$ with negative slope  (again consistent with Fig. 1).
This theoretical curve does exhibit a maximum at
$x\sim 0.18$,
$W\sim 0.428$. Intriguingly, however, it tends to
underestimate the real-history winnings suggesting that the
actual population is exhibiting an additional degree of
co-operation (correlation). A fuller theory
of this mixed system is therefore needed which will
incorporate such additional inter-agent correlations.

In summary, we have studied the performance of a binary
population containing agents with dissimilar memories. Both
the strategy-space (`configurational') and history-space
(`temporal') correlations play a crucial role in the system's
evolution. The  importance of memory has been demonstrated.
The present results may  contribute to a better understanding
of optimization in generic multi-agent systems where resources
are limited, e.g. social dilemmas and Internet congestion
\cite{huberman,arthur,johnson1}.   Assuming that the present
system provides a simple model of a marketplace, our results
suggest that a surprisingly large number of `superior' ($m'=6$)
traders can consistently `beat the market' by uncovering and
exploiting hidden correlations in the system's recent history
{\em without} being directly detrimental to the remainder:
needless to say that the usefulness of such technical
analysis, or `charting', is still a hotly debated issue in the
financial world. 

We thank A. Short for additional computer simulations, and D.
Challet, D. Leonard, D. Sherrington and A. Cavagna for
discussions regarding the pure-population minority game.

\newpage \centerline{\bf Figure Captions}

\bigskip

\noindent Figure 1: Average winnings per agent per turn $W$
for a mixed population of 
$N=101$ players, with memory $m=3$ or $m'=6$, obtained
numerically. Each agent has
$s=7$ strategies.  Each data-point corresponds to an average
over 32 runs. See text for an explanation of `real history'
and `random history'.

\bigskip

\noindent Figure 2:  Average winnings per turn for $m=3$ and
$m'=6$ agents separately. Other parameters are as in Figure 1.
See text for an explanation of `coin-toss'.

\end{document}